
------------------------------ Start of body part 2

%
%
%
%
%
%

\magnification=1200
\tolerance=2000
\hbadness=2000
\overfullrule=0pt


\font\gross=cmbx10 scaled \magstep2
\font\mittel=cmbx10 scaled\magstep1

\font\pl=cmssq8 scaled \magstep1
\font\sc=cmcsc10

\def\RR{{\rm I\!R}}

\def\h#1{{\cal #1}}
\def\a{\alpha}
\def\b{\beta}
\def\g{\gamma}
\def\d{\delta}
\def\eps{\varepsilon}
\def\l{\lambda}
\def\m{\mu}
\def\n{\nu}

\def\om{\omega}
\def\na{\nabla}

\def\Tr{{\rm Tr\,}}

\def\det{{\rm det\,}}

\def\Det{{\rm Det\,}}

\def\sinh{{\rm sinh\,}}
\def\cosh{{\rm cosh\,}}
\def\coth{{\rm coth\,}}
\def\tanh{{\rm tanh\,}}
\def\log{{\rm log\,}}

\def\R{{\cal R}}
\def\F{{\cal F}}
\def\const{{\rm const}}

\def\sq{\Square}
\def\Square{\mathchoice{\square{6pt}}{\square{5pt}}{\square{4pt}}
    {\square{3pt}}}
\def\square#1{\mathop{\mkern0.5\thinmuskip\vbox{\hrule\hbox{\vrule
    \hskip#1 \vrule height#1 width 0pt \vrule}\hrule}\mkern0.5\thinmuskip}}
\def\indunt#1#2#3 {{ #1_{_{_{\hskip -8pt
    \if #30 {\scriptstyle (#2)}
    \else {\scriptstyle (#2_{#3})}\fi }}}}}

\def\frac#1#2{{\textstyle {#1\over #2}}}

\def\ltextindent#1{\hbox to \hangindent{#1\hss}\ignorespaces}
\def\up#1{\leavevmode \raise.16ex\hbox{#1}}

\def\today{\ifcase\month\or January\or February\or March\or April\or May
    \or June\or July\or August\or September\or October \or November
    \or December \fi\space\number\day, \number\year}
\def\heute{\number\day. {\ifcase\month\or Januar\or Februar\or M\"arz
    \or April\or Mai\or Juni\or Juli\or August\or September\or Oktober
    \or November\or Dezember\fi} \number\year}

\def\rightheadline{\it\title\qquad\hfill\rm\folio}
\def\leftheadline{\rm\folio\hfill\it\qquad\author}
\headline={\ifnum\pageno>1{\ifodd\pageno\rightheadline\else\leftheadline\fi}
\else\fi}

\def\author{Ivan G. Avramidi}
\def\title{Heat kernel in quantum field theory}

%

\nopagenumbers

{\null
\vskip-1.5cm
\hskip5cm{ \hrulefill }
\vskip-.55cm
\hskip5cm{ \hrulefill }
\smallskip
\hskip5cm{{\pl \ University of Greifswald (September, 1995)}}
\smallskip
\hskip5cm{ \hrulefill }
\vskip-.55cm
\hskip5cm{ \hrulefill }
\bigskip
\hskip5cm{\ hep-th/9509075}
\bigskip
\hskip5cm{\ to appear in }

\hskip5cm{\ Proceedings of the International Seminar}

\hskip5cm{\ `Quantum Gravity'}

\hskip5cm{\ Moscow, June 19-25, 1995}

\vfill

\centerline{\gross Covariant approximation schemes }
\medskip
\centerline{\gross for calculation of the heat kernel}
\medskip
\centerline{\gross in quantum field theory}
\vskip12pt
\centerline{\sc Ivan G. Avramidi
\footnote{*}{\rm Alexander von Humboldt Fellow}
}\smallskip
\centerline{\sl Department of Mathematics, University of Greifswald}
\centerline{\sl Jahnstr. 15a, 17487 Greifswald, Germany}
\centerline{\sl E-mail: avramidi@math-inf.uni-greifswald.d400.de}
\centerline{\sc and}
\centerline{\sl Research Institute for Physics, Rostov State University}
\centerline{\sl Stachki 194, 344104, Rostov-on-Don, Russia}

\vfill
{\narrower
This paper is an overview on our recent results in the calculation of the heat
kernel in quantum field theory and quantum gravity.
We introduce a deformation of the background fields (including the metric of a
curved spacetime manifold) and study various asymptotic expansions of the heat
kernel diagonal associated with this deformation.
Especial attention is payed to the low-energy
approximation corresponding to the strong slowly varying background
fields.
We develop a new covariant purely algebraic approach for calculating
the heat kernel diagonal in low-energy approximation by taking into account a
finite number
of low-order covariant derivatives of the background fields, and
neglecting all covariant derivatives of higher orders.
Then there exist a set of covariant differential operators that together
with the background fields and their low-order derivatives generate
a finite dimensional Lie algebra.
In the zeroth order of the low-energy perturbation theory, determined by
covariantly constant background, we use this
algebraic structure to present the heat operator
in the form of an average over the corresponding Lie group. This simplifies
considerably the calculations
and allows to obtain closed explicitly covariant formulas for the
heat kernel diagonal.
These formulas serve as the generating functions for the whole sequence
of the Hadamard-Minakshisundaram- De Witt-Seeley coefficients in the low-energy
approximation.}}

\eject

\leftline{\mittel 1. Introduction}
\bigskip

The heat kernel for an elliptic differential operator $H$ acting on sections of
a vector bundle over a manifold $M$ plays a very important role in various
areas of mathematical physics, especially in quantum field theory and quantum
gravity [1-12].
It is defined as the kernel of the one-parameter semigroup (or heat operator),
$U(t)=\exp(-t H)$, viz.
$$
U(t|x,x') = \exp(-t H)\h P(x,x')g^{-1/2}\d(x,x'),\eqno(1.1)
$$
where $\h P(x,x')$ is the parallel displacement
operator of quantum fields $\varphi(x)$ (sections of the vector bundle) from
the point $x$ to the point $x'$  along the geodesic.

 The heat kernel determines among others such fundamental objects of the
quantum field theory as the Green function, the kernel of the resolvent,
$(H+\l)^{-1}$,
the zeta-function, [13]
$$
\zeta(p)=\Tr H^{-p}={1\over\Gamma(p)}\int\limits_0^\infty dt t^{p-1}\Tr U(t),
\eqno(1.2)
$$
the functional determinant, $\Det H$,
and, hence, the one-loop effective action
$$
\Gamma_{(1)}={\frac 1 2}\log\Det H = -{\frac 1 2} \zeta'(0).
\eqno(1.3)
$$
The functional trace `Tr' in (1.2) is defined according to
$$
{\rm Tr}U(t)=\int\limits_M dx g^{1/2} {\rm tr} [U(t)],
\eqno(1.4)
$$
where `tr' is the usual matrix trace and
$$
[U(t)]=U(t|x,x)
\eqno(1.5)
$$
is the heat kernel in coinciding points, so-called  heat kernel diagonal.

In quantum field theory, the manifold $M$ is, usually, taken to be a
$d$-dimensional Riemannian manifold with a metric, $g_{\mu\nu}$, of Euclidean
(positive) signature.
The most important operators are the second order elliptic operators of Laplace
type
$$
H=-\sq + Q,
\eqno(1.6)
$$
where $\sq=g^{\mu\nu}\na_\mu\na_\nu$ is the Laplacian, $\na_\mu$ is a
connection on the vector bundle and $Q$ is an endomorphism of this bundle.
In other words the operator $H$ acts on quantum fields $\varphi(x)$, $Q(x)$ is
a matrix valued potential term, $\na_\mu$ is the covariant derivative defined
with a connection, a Yang-Mills gauge field, ${\cal A}_\mu$. The gauge field
strength (Yang-Mills curvature), $\R_{\m\n}$, is given by the commutator of
covariant derivatives
$$
[\na_\mu,\na_\nu]\varphi={\cal R}_{\mu\nu}\varphi.
\eqno(1.7)
$$

Obviously, the heat kernel is calculable {\it exactly} only in exceptional
cases of background fields configurations,
(see, for example [14]).
On the other hand, to get the quantum amplitudes one has to calculate the
effective action as the
functional of background fields of {\it general} type.
That is why one needs to develop {\it approximate} methods for calculation of
the heat kernel in {\it general} case.

In quantum gravity and gauge theories the effective action is a
{\it covariant} functional, i.e. it is invariant under diffeomorphisms and
local gauge transformations. That is why the approximations for
calculating the effective action must be {\it manifestly covariant},
i.e. they have to preserve the general
covariance at {\it each order}.

Except for the well-known Schwinger-De Witt expansion [1-6] there are two
covariant approximation schemes available [3]: {\it i)} the high-energy one,
which corresponds to weak rapidly varying background fields (short waves), and
{\it ii)} the low-energy approximation corresponding to the strong slowly
varying background fields (long waves). The high-energy approximation was
studied in [5,6,15-17] where the heat kernel and the effective action in
second [5,6,15,16] and third [17] order in background fields (curvatures) were
calculated. The low-energy approximation in various settings was studied in
[18-21]. The authors of these papers summed up some particular terms in the
heat kernel asymptotic expansion, such as the scalar curvature terms [18,19] or
terms without derivatives of the potential
term [20,21] etc.

In our recent
papers [22-27] we studied the low-energy approximation in quantum gravity and
gauge theories and developed a new purely algebraic {\it covariant} approach
for calculating the heat kernel near diagonal. The point is that in low-energy
approximation the covariant derivatives of the curvatures and the potential
term (but not the curvature and the potential term themselves!) are small.
Therefore, one can treat them perturbatively, the zeroth order
of this perturbation theory corresponding to the covariantly constant
background fields.

In particular, the following cases were considered:
\item{i)} covariantly constant gauge field strength and the potential term in
flat space, $\na_\m\R_{\a\b}=\na_\m Q=R_{\a\b\g\d}=0$, [22,23,26],
\item{ii)} covariantly constant Riemann curvature and the potential term
without the Yang-Mills curvature, $\na_\m R_{\a\b\g\d}=\na_\m Q=\R_{\m\n}=0$,
[22, 24-26],
\item{iii)} covariantly constant Yang-Mills curvature and the potential term
with nonvanishing first and second derivatives in flat space, $\na_\m
\R_{\a\b}=\na_\m\na_\n\na_\l Q=R_{\a\b\g\d}=0$ [27].
In the [28] this method was applied for the calculation of the effective
potential and the investigation of the vacuum structure of non-Abelian gauge
theories.

One should stress from the very beginning that our analysis is {\it purely
local.}
We are not interested in the influence of topology but concentrate our
attention rather on local effects. Of course, there are always special global
effects (Casimir like effects, influence of boundaries, presence of closed
geodesics etc.) that do not show up in the local study of the heat kernel.
However, our aim is to investigate only those general physical situations where
the contribution of these effects is small in comparison with local part.  We
are not going to present some exact result for specific background fields, but
to develop some general approximation schemes of calculations. The algebraic
approach elaborated in our papers [22-27] should be thought as a framework for
a perturbation theory in non-homogeneity.

\bigskip
\bigskip
\leftline{\mittel 2. Asymptotic expansions}
\nobreak
\bigskip
\nobreak

Let us call the Riemann curvature tensor $R_{\m\nu\a\b}$, the Yang-Mills
curvature ${\cal R}_{\m\nu}$ and the
potential term $Q$ the {\it background curvatures} or simply curvatures and
denote them
symbolic by $\Re=\{R_{\m\nu\a\b}, {\cal R}_{\m\nu}, Q \}$.
Let us introduce, in addition, the {\it infinite} set of all covariant
derivatives of the curvatures, so-called {\it background jets},
$$
\h J=\{\Re_{(i)}\}, \qquad
\Re_{(i)}=\{\na_{\m_1}\cdots\na_{\m_i}\Re\}.
\eqno(2.1)
$$
The whole set of the jets $\h J$ completely describes the background, at least
locally.

Let us make a deformation of the background fields by introducing some
deformation parameters $\a$ and $\eps$
$$
g_{\m\n} \to g_{\m\n}(\a,\eps), \qquad
\h A_{\m} \to \h A_{\m}(\a,\eps), \qquad
Q \to Q(\a,\eps)
\eqno(2.2)
$$
in such a way that the jets transform uniformly
$$
\Re_{(i)}\to \a\eps^i\Re_{(i)}.
\eqno(2.3)
$$
Such deformation of the background fields lead to the corresponding deformation
of the operator $H$ (1.6)
$$
H\to H(\a,\eps)
\eqno(2.4)
$$
and the heat kernel
$$
U(t) \to U(t;\a,\eps).
\eqno(2.5)
$$

Let us note that because of the transformation law (2.3) this deformation is
{\it manifestly covariant}. Therefore, it gives a natural framework to develop
various approximation schemes based on asymptotic expansions of the heat kernel
in the the deformation parameters.
It is obvious that the limit $\a\to 0$ corresponds to the small curvatures,
$\Re\to 0$, i.e. to the covariant perturbation theory [17], and the other
limit, $\eps\to 0$, corresponds to small covariant derivatives of the
curvatures, $\na_\m \Re\to 0$, i.e. to the long-wave approximation [22-27].
More precisely, we recognize two cases:

\noindent
i) the high-energy approximation,
$$
\na\na\Re\gg\Re\Re \qquad {\rm or} \qquad \eps^2 \gg \a,
$$
and ii) the low-energy approximation,
$$
\na\na\Re\ll\Re\Re \qquad {\rm or} \qquad \eps^2 \ll \a.
$$

\bigskip
\leftline{\bf 2.1. Schwinger - De Witt asymptotic expansion}
\nobreak\bigskip\nobreak

First of all, there is an asymptotic expansion of the heat kernel as $t\to 0$
(Schwinger - De Witt expansion)
[2,5-10]
$$
\eqalignno{
& [U(t;\a,\eps)] \sim (4\pi t)^{-d/2}
\sum\limits_{k\ge 0} {(-t)^k\over k!} a_k(\a,\eps).        &(2.6)\cr}
$$

This expansion is purely local  and does not depend, in fact,
on the global structure of the manifold. Its famous coefficients
$a_k$, Hadamard - Minakshisundaram - De Witt - Seeley (HMDS) coefficients,
are  local invariants built from the background curvatures
and their covariant derivatives
[1-12,29-33].
The HMDS-coefficients play a very important role both in physics and
mathematics
and are closely connected with various sections of mathematical
physics such as spectral geometry, index theorem, trace anomalies, Korteweg -
de~Vries
hierarchy etc.
[7,12,33].

One can classify all the terms in $a_k$ according to the number of
curvatures and their derivatives.
First, there are terms linear in the curvature, then it follows the class of
terms quadratic in the curvature, etc.. The last class of terms does not
contain any
covariant derivatives at all but only the powers of the curvatures.
This general structure emerges by the expansion of $a_k$ in the deformation
parameters
$$
a_k(\a,\eps)=\sum_{0\le n\le k}\a^n\eps^{2k-2n}a_{k,n}.
\eqno(2.7)
$$
Here $a_{k,n}$ are the homogeneous parts of $a_k$ of order $n$  in the
curvatures
that can be symbolically written in the form
$$
a_{k,n}=\sum_{i_1,\dots,i_n \ge 0\atop i_1+\cdots+i_n=k-2n}
\sum \Re_{(i_1)} \cdots \Re_{(i_n)},
\eqno(2.8)
$$
where the second summation is over different invariant structures.
The first coefficient reads simply
$$
a_0=1,
\eqno(2.9)
$$
and the higher order coefficients $a_k$, $(k\ge 1)$ have the following
homogeneous parts [5,6]
$$
\eqalignno{
a_{k,0}&=0,&\cr
a_{k,1}&=-\a^{(1)}_k \sq^{k-1}Q+\a^{(1)}_k\sq^{k-1} R, &\cr
a_{k,2}&=
\b^{(1)}_k Q\sq^{k-2}Q
+2\b^{(2)}_k \h R_{\alpha\mu}\nabla^\alpha\sq^{k-3}
\nabla_\nu\h R^{\nu\mu}
-2\b^{(3)}_k Q\sq^{k-2}R &\cr
&+\b^{(4)}_k R_{\mu\nu}\sq^{k-2}
R^{\mu\nu}+\b^{(5)}_k R\sq^{k-2}R
+\na\left(\sum_{0\le i\le 2k-3}\sum\na^i\Re\na^{2k-3-i}\Re\right),&(2.10)\cr
& \dots &\cr
a_{k,k}&=\sum\Re^k, &\cr}
$$
where $\a^{(i)}_k$ and $\b^{(i)}_k$ are numerical constants. Note that
altogether there are only five quadratic invariant structures (up to the total
derivatives) but very many structures of the type $\Re^k$.

The first coefficients, $a_0, a_1, a_2$, were
calculated a long time ago by De Witt
[2], $a_3$ was calculated by Gilkey
[29] and
the next coefficient, $a_4$, was calculated for the first time in general case
in our PhD thesis [5] and published in
[6,30,31]
and in the case of scalar operators in
[32].
The linear and quadratic parts in the HMDS-coefficients, i.e. $a_{k,1}$ and
$a_{k,2}, (k\ge 2)$, were also calculated in our PhD thesis [5] and published
in [6,15,16]. The quadratic part was calculated only up to a total derivative.
The same results were obtained completely independent in [34]. The next cubic
order in curvature, $a_{k,3}$ was studied in [17]. The terms without the
derivatives, $a_{k,k}$, in general case are unknown. The calculation of these
terms is an open and very interesting and important problem.

\bigskip
\leftline{\bf 2.2. High-energy asymptotic expansion}
\nobreak
\bigskip
\nobreak

Let us consider now the asymptotic expansion in the limit $\a\to 0$ of the
perturbation theory. One can show that it has the form
$$
\eqalignno{
& [U(t;\a,\eps)] \sim (4\pi t)^{-d/2}
\sum\limits_{n\ge 0} (\a t)^nh_n(t;\eps),      &(2.11)\cr}
$$
where $h_n(t,\eps)$ are some {\it nonlocal} functionals that have the following
asymptotic expansion as $t\to 0$
$$
h_n(t;\eps)\sim \sum_{l\ge 0} {(-1)^{n+l}\over (n+l)!}(\eps^2 t)^l a_{n+l,n}.
\eqno(2.12)
$$
The first functionals $h_n$ are [5,6,15,16]
$$
\eqalignno{
h_0(t;\eps)&=1, &\cr
h_1(t;\eps)&=t\left\{F_1(\eps^2 t\sq)Q-F_2(\eps^2 t\sq)R\right\},&(2.13)\cr
h_2(t;\eps)&={t^2\over 2}\biggl\{QF_{(1)}(\eps^2t\Square)Q
+2\h R_{\alpha\mu}\nabla^\alpha{{1}\over {\Square}}F_{(3)}
(\eps^2t\Square)\nabla_\nu\h R^{\nu\mu}
-2QF_{(2)}(\eps^2t\Square)R                                   &\cr
&+R_{\mu\nu}F_{(4)}(\eps^2t\Square)R^{\mu\nu}
+RF_{(5)}(\eps^2t\Square)R\biggr\}+ {\rm total \ derivative},&\cr}
$$
where $F_{(i)}(z)$ are the formfactors, i.e. some analytic functions. One can
show that the formfactors $F_{(i)}(z)$ are {\it entire} functions, i.e. they
are analytic in the whole complex plane. The explicit form of these functions
was obtained in our PhD thesis [5] and published in the papers [6,15,16].
The {\it third} order in curvatures of the covariant perturbation theory was
investigated in [17].

\bigskip
\leftline{\bf 2.3. Low-energy asymptotic expansion}
\bigskip

The low-energy approximation corresponds to the asymptotic expansion of the
deformed heat kernel as $\eps\to 0$
$$
\eqalignno{
& [U(t;\a,\eps)] \sim (4\pi t)^{-d/2}
\sum\limits_{l\ge 0} (\eps^2 t)^l u_l(t;\a).      &(2.14)\cr}
$$
The coefficients $u_l$ are essentially {\it non-perturbative} functionals that
have the following perturbative asymptotic expansion as $t\to 0$
$$
u_l(t;\a)\sim \sum_{n\ge 0}{(-1)^{n+l}\over(n+l)!}(\a t)^na_{l+n,n}.
\eqno(2.15)
$$
The zeroth order of this approximation,
$$
[U(t;\a,\eps)]\Big\vert_{\eps=0}\sim (4\pi t)^{-d/2}u_0(t;\a),
\eqno(2.16)
$$
corresponds to covariantly constant background
$$
\Re_{(i)}=0 \qquad {\rm for} \ i\ge 1,
$$
or, more explicitly,
$$
\na_\m R_{\a\b\g\d} = 0,\qquad \na_\m{\cal R}_{\a\b}=0,\qquad \na_\m Q = 0.
							\eqno(2.17)
$$
The zeroth order functional $u_0(t;\a)$ has the following perturbative
asymptotic expansion
$$
u_0(t;\a)\sim\sum_{n\ge 0}{(-1)^n\over n!}(\a t)^na_{n,n},
\eqno(2.18)
$$
or, symbolically,
$$
u_0(t;\a)\sim\sum_{n\ge 0}\sum(\a t\Re)^n,
\eqno(2.19)
$$
and can be viewed on as the {\it generating function} for that part of the
HMDS-coefficients, $a_{k,k}$, that does not contain any covariant derivatives
(last eq. in (2.10)).


\bigskip
\bigskip
\leftline{\mittel 3. Algebraic approach}
\bigskip

There exist a very elegant indirect possibility to construct the heat kernel
 without solving the heat equation but using only the commutation
relations of some  covariant first order differential operators [22-27].
The main idea is in a generalization of the usual Fourier transform to the
case of operators and consists in the following.
Let us consider for a moment a trivial case of vanishing curvatures but not
the potential term
$$
R_{\a\b\g\d} = 0,\qquad {\cal R}_{\a\b}=0,\qquad Q \ne 0.
							\eqno(3.1)
$$
In this case the operators of covariant derivatives obviously commute and
form together with the potential term an Abelian algebra
$$
[\na_\m,\na_\nu]=0,\qquad [\na_\m, Q]=0. \eqno(3.2)
$$
It is easy to show that the heat {\it operator} can be presented
in the form
$$
\exp(t\sq)=(4\pi t)^{-d/2}\int dk g^{1/2}
\exp\left(-{1\over 4t}k^\m g_{\m\nu}k^\nu\right)
\exp\left(k^\m \na_\m\right),
\eqno(3.3)
$$
where it is assumed that the covariant derivatives commute
also with the metric $[\na_\m,g_{\a\b}]=0$.
Acting with this operator on the $\d$-function and using the obvious
relation
$$
\exp(k^\m \na_\m)\d(x,x')\Big\vert_{x=x'}=\d(k)
\eqno(3.4)
$$
one can simply integrate over $k$ in (3.3) to obtain the heat kernel in
coordinate representation.
The heat kernel diagonal is given then by
$$
[U(t)]=(4\pi t)^{-d/2}\exp\left(-tQ\right). \eqno(3.5)
$$

In fact, the covariant differential operators $\na$ do not commute and the
commutators of
them are proportional to the curvatures $\Re$.
The commutators of covariant derivatives with the curvatures give the first
derivatives of the curvatures , i.e. the jets $\Re_{(1)}$, the commutators of
covariant derivatives with $\Re_{(1)}$ give the second jets $\Re_{(2)}$ etc.
$$
\eqalignno{
[\na, \na]&=\Re, &\cr
[\na, \Re]&=\Re_{(1)},&\cr
&\dots &\cr
[\na, \Re_{(i)}]&=\Re_{(i+1)},&(3.6)\cr
&\dots &\cr}
$$
The commutators of jets themselves are the product of jets again
$$
[\Re_{(i)}, \Re_{(k)}]=\Re_{(i+k+2)}+\sum_{0\le n\le
k}\sum\Re_{(n)}\Re_{(i+k-n)},
\eqno(3.7)
$$
(in Abelian case all such commutators vanish).

Thus the operators of covariant differentiation $\na$ together with the whole
set of the jets $\h J$ form an {\it infinite} dimensional Lie algebra $\h
G=\{\na, \Re_{(i)}\}$.
To gain greater insight into how the low-energy heat kernel looks like, one can
take into account a {\it finite} number of low-order jets, i.e. the low-order
covariant derivatives of the background fields, and neglect all the higher
order jets, i.e. the covariant derivatives of higher orders.
Then one can show that there exist a set of covariant differential operators
that together with the background fields and their low-order derivatives
generate a {\it finite} dimensional Lie algebra $\h G'$.
This procedure is very similar to the polynomial approximation of functions of
real variables. The difference is that we are dealing, in general, with the
{\it covariant} derivatives and the curvatures.

Thus one can try to generalize
the above idea in such a way that (3.3) would be the zeroth approximation
in the commutators of the covariant derivatives, i.e. in the curvatures.
Roughly speaking, we are going to find a representation of the heat kernel
{\it operator}
in the form
$$
\exp(t\sq) = \int dk \Omega(t,k)
\exp\left\{-{1\over 4t}k^A\Pi_{AB}(t)k^B\right\}
\exp(k^A\xi_A),
\eqno(3.8)
$$
where $\xi_A=\{X_a, Y_i\}$, $X_a=X^\m_a(x)\na_\m$ are some first order
differential operators and $Y_i(x)$ are some functions. The functions $\Pi(t)$
and $\Omega(t,k)$ are expressed in terms of
the commutators of this operators, i.e. in terms of the curvatures.

In general, the operators $\xi_A$ do not form a closed finite dimensional
Lie algebra because at each stage taking more commutators there appear more
and more derivatives of the curvatures. If one restricts oneself to the
low-order jets, this algebra closes and becomes finite dimensional.

Using this representation one could, as above, act with $\exp(k_A\xi^A)$
on the $\d$-function on $M$ to get the heat kernel. The main point of
this idea is that it is much easier to calculate the action of
the exponential of the {\it first} order operator $k^A\xi_A$ on the
$\d$-function than that of the exponential of the second order operator
$\sq$.

\bigskip
\bigskip
\leftline{\mittel 4. Heat kernel in flat space}
\bigskip

In this section we consider the
more complicated case of nonvanishing covariantly constant Yang-Mills curvature
in the flat space
$$
R_{\a\b\g\d} = 0,\qquad \na_\m\h R_{\a\b}=0.
							\eqno(4.1)
$$
As we will study only {\it local} effects in the low-energy approximation, we
will not take care about the topology of the manifold $M$. To be precise one
can take, for example, $\RR^d$.

\bigskip
\leftline{\bf 4.1. Covariantly constant potential term}
\bigskip

First we consider the case of covariantly constant potential term
$$
\na_\m Q=0.
\eqno(4.2)
$$
In this case the covariant derivatives
form a {\it nilpotent} Lie algebra
$$
\eqalignno{
&[\na_\m,\na_\nu]=\h R_{\m\nu}, &(4.3)\cr
&[\na_\m,\h R_{\a\b}]=[\na_\m,Q]=[\h R_{\m\nu},\h R_{\a\b}]=[\h
R_{\m\nu},Q]=0.\cr}
$$

For this algebra one can prove a theorem expressing the heat
operator in terms of an average over the corresponding Lie group [22,23]
$$
\eqalignno{
\exp(t\sq) =& (4\pi t)^{-d/2}
\det\left({t\h R\over \sinh(t\h R)}\right)^{1/2}&\cr
&\int dk g^{1/2}
\exp\left\{-{1\over 4t}k_\l(t\h R \coth(t\h R))^\l_{\ \nu}
k^\nu\right\}
\exp\left(k^\m \na_\m\right),	&(4.4)\cr}
$$
where $\h R=\{\h R^\m_{\ \nu}\}$ means the matrix with spacetime indices
and the determinant
is taken with respect to these indices, other indices being
intact.

It is not
difficult to show that [22,23]
$$
\exp(k^\m \na_\m)\h P(x,x')\d(x,x')\Big\vert_{x=x'}
=\d(k).
\eqno(4.5)
$$
Subsequently, the integral over $k^\m$ becomes trivial and one obtains
immediately the heat kernel diagonal
$$
[U(t)]=(4\pi t)^{-d/2}
\det\left({t\h R \over \sinh(t\h R)}\right)^{1/2}
\exp\left(-tQ\right). \eqno(4.6)
$$

Expanding it in a power series in $t$ one can
find {\it all} coefficients $a_{k,k}$ (2.10), i.e. {\it all} covariantly
constant terms in {\it all} HMDS-coefficients $a_k$ (2.7).

As we have seen the contribution of the Yang-Mills curvature
is not as trivial as that of the potential term. However, the algebraic
approach does work in this case too. This is the generalization of the
well known Schwinger result [1] in quantum electrodynamics. It is a good
example how one can get the heat
kernel without solving any differential equations but using only the
algebraic properties of the covariant derivatives.
This result was applied for calculating the one-loop low-energy effective
action in the non-Abelian gauge theory and for studying the stability of
the vacuum [28].


\bigskip
\bigskip
\leftline{\bf 4.2. Inclusion of the first and second derivatives of the
potential term}
\bigskip

 Now we consider the case when the first and the second derivatives of the
potential term do not vanish but all the higher derivatives do, i.e
$$
\na_\mu\na_\nu\na_\l Q=0.
\eqno(4.7)
$$
Besides we assume the background to be {\it Abelian}, i.e. all the nonvanishing
background quantities, ${\cal R}_{\a\b}$, $Q$, $Q_{;\mu}$, $Q_{;\nu\mu}$,
commute with each other. Thus we have a nilpotent Lie algebra $\{\na_\m, {\cal
R}_{\a\b}, Q, Q_{;\mu}, Q_{;\nu\mu}\}$
$$
[\na_\mu, \na_\nu]={\cal R}_{\mu\nu},
$$
$$
[\na_\m,Q]=Q_{;\m}
$$
$$
[\na_\m,Q_{\n}]=Q_{;\n\m}
\eqno(4.8)
$$
$$
[{\cal R}_{\a\b}, {\cal R}_{\mu\nu}]
=[{\cal R}_{\a\b}, Q_{;\nu}]
=[{\cal R}_{\a\b}, Q_{;\mu\nu}]
=[Q, Q_{;\mu}]
=[Q, Q_{;\mu\nu}]=
[Q_{;\mu}, Q_{;\a\b}]=0,
$$
where $Q_{;\mu}\equiv\na_\mu Q$, $Q_{;\nu\mu}\equiv\na_\mu\na_\nu Q$.

For our purposes, it is helpful to introduce the following parametrization of
the potential term
$$
Q=M-\b^{ik}L_iL_k,
\eqno(4.9)
$$
where $(i=1,\dots,p)$, $p\le d$, $\b^{ik}$ is some constant symmetric
nondegenerate $p\times p$ matrix, $M$ is a covariantly constant matrix and
$L_i$ are some matrices with vanishing {\it second} covariant derivative
$$
\na_\m M=0, \qquad \na_\m\na_\n L_i=0.
\eqno(4.10)
$$

This gives us another nilpotent Lie algebra, $\{\na_\mu,$ ${\cal R}_{\a\b},$
$M,$ $L_i,$ $L_{i;\m}\}$, with following nontrivial commutators
$$
[\na_\mu, \na_\nu]={\cal R}_{\mu\nu}, \qquad
[\na_\mu, L_i] = L_{i;\m},
\eqno(4.11)
$$
and the center $\{{\cal R}_{\a\b},$ $M,$ $L_i,$ $L_{i;\mu}\}$.
Introducing the generators $\xi_A=(\na_\m, L_i)$, $(A=1,\dots, D)$, $D=d+p$,
one can rewrite these commutation relations in a more compact form
$$
[\xi_A, \xi_B]=\F_{AB}, \qquad
\eqno(4.12)
$$
$$
[\xi_A, \F_{CD}]=[\F_{AB}, \F_{CD}]=0,
$$
where $\F_{AB}$ is a matrix
$$
(\F_{AB})=\left(\matrix{
\R_{\m\n} & L_{i;\m}\cr
-L_{k;\n} & 0       \cr}\right),
\eqno(4.13)
$$
that we call the {\it generalized} curvature.
The operator $H$ (1.4) can now be written in the form
$$
H=-\g^{AB}\xi_A \xi_B+M,
\eqno(4.14)
$$
where
$$
(\g^{AB})=\left(\matrix{g^{\m\n} & 0 \cr
			0 & \b^{ik} \cr}\right).
\eqno(4.15)
$$
The matrices $\b^{ik}$ and $\g^{AB}$ play the role of metrics and can be used
to raise and to lower the small and the capital Latin indices respectively.

Note that the algebra (2.10) is essentially of the same type as (4.3).
 For algebras of this kind the heat operator is given by the integral over the
corresponding Lie group [22,23]
$$
\eqalignno{
\exp(-t H)=&(4\pi t)^{-D/2}
\det\left({\sinh(t\F)\over t\F}\right)^{-1/2}\exp(-t M)
&\cr&\times\int\limits_{\RR^D} d k \g^{1/2}
\exp\left\{-{1\over 4t}k_A(t\F \coth(t\F))^A_{\ B}k^B\right\}
\exp(k^A \xi_A),&(4.16)\cr}
$$
where $\g=\det\g_{AB}$.

Thus we have expressed the heat kernel operator in terms of the operator
$\exp(k^A \xi_A)$. The integration over $k$ in (4.16) is Gaussian except for
noncommutative part.
Splitting the integration variables $(k^A)=(q^\m, \om^i)$ and using the
Campbell-Hausdorf formula we obtain [27]
$$
\exp(k^A\xi_A)\d(x,x')\Big\vert_{x=x'}=\exp(\om^i L_i)\d(q),
\eqno(4.17)
$$
hence taking off the integration over $q$. After integrating over $\om$  we
obtain the heat kernel diagonal in a very simple form [27]
$$
[U(t)]=(4\pi t)^{-d/2}\Phi(t)
\exp\left\{-t Q + \frac 1 4 t^3 Q_{;\m}\Psi^{\m\n}(t)Q_{;\n}\right\}.
\eqno(4.18)
$$
where
$$
\Phi(t)=\det\left({\sinh(t\F)\over t\F}\right)^{-1/2}
\det(1+t^2C(t)P)^{-1/2},
\eqno(4.19)
$$
$$
\Psi(t)=(\Psi^\m_{\ \n}(t))=(1+t^2C(t)P)^{-1}C(t),
\eqno(4.20)
$$
$P$ is the matrix of second derivatives of the potential term,
$$
P=\left\{\frac 1 2 Q^{;\m}_{\ ;\n}\right\}
\eqno(4.21)
$$
 and
the matrix $C(t)=\{C^\m_{\ \n}(t)\}$ is defined by
$$
C(t)=\oint\limits_C{dz\over 2\pi i}t \coth({tz^{-1}})(1-z\h R-z^2 P)^{-1}.
\eqno(4.22)
$$

 The formula (4.18) exhibits the general structure of the heat kernel diagonal.
Namely, one sees immediately how the potential term and its first derivatives
enter the result. The complete nontrivial information is contained only in a
scalar, $\Phi(t)$, and a tensor, $\Psi_{\m\n}(t)$, functions which are
constructed purely from the Yang-Mills curvature $\R_{\m\n}$ and the {\it
second} derivatives of the potential term, $\na_\m\na_\n Q$. So we conclude
that the coefficients of the heat kernel asymptotic expansion are constructed
from three different types of scalar (connected) blocks, $Q$, $\Phi_{(n)}(\R,
\na\na Q)$ and $\na_\m Q\Psi^{\m\n}_{(n)}(\R, \na\na Q)\na_\n Q$.

In a special case, when the matrices $\R$ and $P$ commute, i.e.
$$
[\R, P]=0, \qquad {\rm or} \qquad \R^\m_{\ \n}P^\n_{\ \a}=P^\m_{\ \n}\R^\n_{\
\a},
\eqno(4.23)
$$
the heat kernel diagonal simplifies considerably [27]
$$
\eqalignno{
[U(t)]&=(4\pi t)^{-d/2}\det\left(\sinh(t\Delta)\over t\Delta\right)^{-1/2}&\cr
&\times\exp\left\{-t Q + \frac 1 4 t Q_{;\m}
\left[{1\over P}
\left({\Delta\over 2tP}{\cosh(t\R)-\cosh(t\Delta)\over \sinh(t\Delta)}
+1\right)\right]^\m_{\ \n}
Q^{;\n}\right\}, &(4.24)\cr}
$$
where
$$
\Delta=\sqrt{\R^2+4P}.
\eqno(4.25)
$$
If the second derivatives of the potential vanish, $P_{\m\n}=\frac 1 2
\na_\m\na_\n Q=0$, then we have therefrom
$$
\eqalignno{
[U(t)]=&(4\pi t)^{-d/2}\det\left(\sinh(t\R)\over t\R\right)^{-1/2}&\cr
&\times
\exp\left\{-t Q + \frac 1 4 t Q_{;\m}
\left({{1\over \R^2}(t\R\coth(t\R)-1)}\right)^\m_{\ \n}
Q^{;\n}\right\}. &(4.26)\cr}
$$
This is the first order correction to the case of covariantly constant
potential [22] when additionally the {\it first} derivatives of the potential
are taken into account.

In the case of vanishing Yang-Mills curvature, $\R=0$, we have similarly
$$
\eqalignno{
[U(t)]=&(4\pi t)^{-d/2}\det\left(\sinh(2t\sqrt P)\over 2t\sqrt P\right)^{-1/2}
&\cr&
\times\exp\left\{-t Q
- \frac 1 4 Q_{;\m}
\left({\tanh(t\sqrt P)-t\sqrt P\over P^{3/2}}\right)^\m_{\ \n}
Q^{;\n}\right\}. &(4.27)\cr}
$$
This determines the low-energy approximation without the gauge fields.


\bigskip
\bigskip
\leftline{\bf 4.3. Trace of the heat kernel}
\bigskip

Let us now calculate the trace of the heat kernel. We assumed that the
background fields satisfy the low-energy conditions (4.7) in some region of the
manifold $M$. Let us suppose the manifold $M$ to be $\RR^d$ and the conditions
(4.7) to hold everywhere. Then the formula for the heat kernel diagonal (4.18)
is valid everywhere too. Let the matrix $P$ (4.21) to be nondegenerate, then
one can integrate (4.18) over $\RR^d$ to get [27]
$$
\Tr U(t)=(2t)^{-d}\det\left(\sinh(t\F)\over t\F\right)^{-1/2}\det
P^{-1/2}\exp(-tM).
\eqno(4.28)
$$
In particular case of commuting matrices $\R$ and $P$ the trace of the heat
kernel takes especially simple form
$$
\Tr U(t)=\det\{2(\cosh(t\Delta)-\cosh(t\R))\}^{-1/2}\exp(-tM),
\eqno(4.29)
$$
which reduces to
$$
\Tr U(t)=\det(2\sinh(t\sqrt P))^{-1}\exp(-tM),
\eqno(4.30)
$$
when $\R=0$.

It should be noted that these expressions have  {\it nonclassic} asymptotics,
$\Tr U(t)\sim\const\cdot t^{-d}$ instead of the usual standard one $\Tr
U(t)\sim\const\cdot t^{-d/2}$ that holds on compact manifolds.
The standard form of the asymptotics of the trace of the heat kernel like
(2.11) is the basis for the regularization and renormalization procedure in
quantum field theory [2]. That is why the non-standard asymptotics may cause
serious technical problems in the theory of quantum fields on noncompact
manifolds with background fields that do not fall off at infinity. For example,
the analytical structure of the zeta function (1.2) in non-standard case will
be completely different. This is the consequence of the fact that in this
non-standard situation the physical quantum states are not well defined.


\bigskip
\bigskip
\leftline{\mittel 5. Heat kernel in symmetric spaces}
\nobreak\bigskip\nobreak

Let us now try to generalize the algebraic approach to the case of the
{\it curved} manifolds with covariantly constant Riemann curvature, covariantly
constant potential and vanishing Yang-Mills curvature,
$$
\na_\m R_{\a\b\g\d}=0, \qquad \h R_{\m\n}=0, \qquad \na_\m Q=0.
\eqno(5.1)
$$

A complete simply connected Riemannian manifold with covariantly constant
curvature is called symmetric space.
A symmetric space is said to be of compact, noncompact {\rm or} Euclidean type
if all sectional curvatures $K(u,v)=R_{abcd}u^av^bu^cv^d$
are positive, negative or zero. A direct product of symmetric spaces of compact
and noncompact types is called {\it semisimple} symmetric space. It is well
known that a generic complete simply connected Riemannian symmetric space is a
direct product of a flat space and a semisimple symmetric space [35,36].

It should be repeated here once more that our analysis in this paper is purely
{\it local}. We are looking for a {\it universal local} function of the
curvature, $u_0(t)$, (2.16) that describes adequately  the low-energy limit of
the heat kernel diagonal. Our minimal requirement is that this function should
reproduce {\it all} the terms without covariant derivatives of the curvature in
the local Schwinger-De Witt asymptotic expansion of the heat kernel (2.6),
(2.18), i.e.
it should give {\it all} the HMDS-coefficients $a_{k}$ (2.7) for {\it any}
symmetric space.

Since the HMDS-coefficients have a {\it universal} explicit structure [7],  it
is obvious that any flat subspaces do not contribute in $a_k$. Moreover, since
HMDS-coefficients $a_k$ are analytic in the curvature it is evident that
to find this universal structure it is sufficient to consider only symmetric
spaces of {\it compact} type with positive curvature.
Using the factorization property of the heat kernel [7] and the duality between
compact and noncompact symmetric spaces [35,36] one can obtain then the results
for the general case by analytical continuation.
That is why  we consider only the case of symmetric spaces of {\it compact}
type.

First of all, we choose a frame $e^\m_a(x,x')$ that is {\it covariantly
constant (parallel)} along the geodesic between the points $x$ and $x'$.
 Let us consider the Riemann tensor in
more detail. It is obvious that the frame components of the curvature
tensor of a symmetric space are constant.
For {\it any} Riemannian
manifold they can be presented in the form
$$
R_{abcd} = \b_{ik}E^i_{\ ab}E^k_{\ cd}, \eqno(5.2)
$$
where $E^i_{ab}$, $(i=1,\dots, p; p \le d(d-1)/2)$, is some set of
antisymmetric matrices and $\b_{ik}$ is some symmetric nondegenerate
$p\times p$ matrix.
The traceless matrices $D_i=\{D^a_{\ ib}\}$ defined by
$$
D^a_{\ ib}=-\b_{ik}E^k_{\ cb}g^{ca}= - D^a_{\ bi} \eqno(5.3)
$$
are known to be the generators of the {\it holonomy algebra} ${\cal H}$ [36]
$$
[D_i, D_k] = F^j_{\ ik} D_j, \eqno(5.4)
$$
where $F^j_{\ ik}$ are the structure constants.

In symmetric spaces there exists a
 much wider Lie algebra
${\cal G}$ of dimension $D=p+d$ [24-26]
$$
[C_A, C_B]=C^C_{\ AB}C_C,             \eqno(5.5)
$$
where the structure constants $C^A_{\ BC}$, $(A=1,\dots, D)$ are defined by
$$
C^i_{\ ab}=E^i_{\ ab}, \quad C^a_{\ ib}=D^a_{\ ib},
\quad C^i_{\ kl}=F^i_{\ kl}, \eqno(5.6)
$$
$$
C^a_{\ bc}=C^i_{\ ka}=C^a_{\ ik}=0,
$$
and $C_A=\{C^B_{\ AC}\}$ are the generators of adjoint representation.
Thus the structure of the algebra ${\cal G}$ is
completely determined by the curvature tensor of symmetric space.

Moreover, in case of semisimple symmetric space the algebra $\h G$ is
isomorphic to the algebra of the
infinitesimal isometries, i.e. the Killing vector fields $\xi_A$, [35,36]
$$
[\xi_A,\xi_B]=C^C_{\ AB}\xi_C.    \eqno(5.7)
$$
Therefore, the curvature tensor of the semisimple symmetric space completely
determines the structure of the group of isometries too.


In semisimple symmetric
spaces the Laplacian  can be presented in terms of generators of isometries
[24-26]
$$
\sq = g^{\m\nu}\na_\m\na_\nu = \g^{AB}\xi_A\xi_B ,         \eqno(5.8)
$$
where
$$
\g^{AB} = \left(\matrix{ g^{ab} & 0             \cr
	       0                & \b^{ik}        \cr}\right) \eqno(5.9)
$$
and $\b^{ik}=(\b_{ik})^{-1}$.

Using this representation one can prove a theorem expressing the heat operator
in terms of some average over the group of isometries $G$ [24-26]
$$
\eqalignno{
\exp(t\sq) = &(4\pi t)^{-D/2} \int d k \g^{1/2}
	\det\left({\sinh(k^AC_A/2)\over k^AC_A/2}\right)^{1/2} &\cr
	& \times\exp\left\{ -{1\over 4t}k^A\g_{AB}k^B
	+ {1\over 6} R_G t\right\}\exp(k^A\xi_A) &(5.10)\cr}
$$
where $\g=\det\g_{AB}$, $\g_{AB}=(\g^{AB})^{-1}$ and $R_G$ is the scalar
curvature of the group of
isometries $G$
$$
R_G= -{1\over 4}\g^{AB} C^C_{\ AD}C^D_{\ BC}. \eqno(5.11)
$$
The proof of this theorem is given in [24,25].

Splitting the integration variables $k^A = (q^a, \om^i)$
one can find first
the action of the isometries on the $\d$-function [24-26]
$$
\exp\left(k^A\xi_A\right)g^{-1/2}\d(x,x')\Big\vert_{x=x'}
=\det\left({\sinh(\om^iD_i/2)\over \om^iD_i/2}\right)^{-1}
\eta^{-1/2}\d(q),  \eqno(5.12)
$$
where $\eta=\det g_{ab}$.
Then one can easily integrate over $q$ in (5.10)
to get heat kernel diagonal [24-26]
$$
\eqalignno{
[U(t)]= &(4\pi t)^{-D/2}\int d\om \b^{1/2}
\det\left({\sinh(\om^iF_i/2)\over \om^iF_i/2}\right)^{1/2}
\det\left({\sinh(\om^iD_i/2)\over \om^iD_i/2}\right)^{-1/2} &\cr
&\times\exp\left\{ - {1\over 4 t}\om^i\b_{ik}\om^k
- \left(Q - {1\over 8} R
- {1\over 6} R_H \right) t \right\},                     &(5.13)\cr}
$$
where $\b=\det \b_{ik}$, $F_i=\{F^k_{\ ij}\}$ are the generators of the
holonomy algebra in adjoint representation and $R_H$ is the scalar
curvature of the holonomy group $H$
$$
R_H = -{1\over 4} \b^{ik} F^m_{\ \ il}F^l_{\ km}. \eqno(5.14)
$$

One can present this result also in an alternative nontrivial rather
{\it formal} way without any integration [24-26]
$$
\eqalignno{
[U(t)](t)&=(4\pi t)^{-d/2}\exp\left\{\left({1\over 8}R
+{1\over 6}R_H-Q\right)t\right\}
\det\left({\sinh(\sqrt t \partial^kF_k/2)\over \sqrt t \partial^kF_k/2}
\right)^{1/2}&\cr
&\times
\det\left({\sinh(\sqrt t \partial^kD_k/2)\over \sqrt t
\partial^kD_k/2}\right)^{-1/2}
\exp\left(p_n\b^{nk}p_k\right)\Bigg\vert_{p=0}. &\cr
&       &(5.15)\cr}
$$
where $p_i$ are some auxiliary variables and $\partial^k=\partial/\partial
p_k$.
This formal solution should be understood as a power series in the
derivatives $\partial^i$ that is well defined and determines the
heat kernel asymptotic expansion at $t\to 0$, i.e. {\it all}
HMDS-coefficients $a_k$.

Let us stress that the closed formulae obtained in this
section are {\it exact} (up to possible nonanalytic topological contributions)
and {\it manifestly covariant} because they are expressed in terms of
the invariants of the holonomy group $H$, i.e. the invariants of the
curvature tensor. They can be used now to generate {\it all}
HMDS-coefficients $a_k$ for {\it any} symmetric space, i.e. for
{\it any space with covariantly constant curvature}, simply by
expanding it in a power series in $t$. Thereby one finds {\it all
covariantly constant terms in {\it all} HMDS-coefficients} in a manifestly
covariant way. This gives a very nontrivial example how the heat kernel
can be constructed using only the commutation relations of some differential
operators, namely the generators of infinitesimal isometries of the
symmetric space. We are going to obtain the explicit formulae in a
further work.

Although we considered for simplicity the case of symmetric space of {\it
compact}
type, i.e. with positive sectional curvatures, i.e. positive definite matrix
$\b_{ik}$,
it is not difficult to generalize our results to the general
case using the duality relation and the {\it analytic continuation}.
This means that our formulae for the asymptotic expansion of the heat kernel
should be valid in
general case of arbitrary symmetric space too. Moreover, they do not depend on
the signature of the spacetime metric and should also
be valid for the case of Lorentzian signature.


\bigskip
\bigskip
\leftline{\mittel 6. Conclusion}
\bigskip

We have presented a brief overview of recent results in
studying the heat kernel obtained in our papers [22-28]. We
discussed some ideas connected with the structure of the asymptotic expansions
of the heat kernel with respect to some deformation parameters. These
asymptotic expansions allow to develop a new scheme for covariant calculation
of the heat kernel in low-energy approximation and to calculate explicitly the
heat kernel diagonal in zeroth order.
The main idea of this approximation scheme is to employ the low-energy
background jet algebra.

We obtained closed formulas for the heat kernel diagonal that can be treated as
a resummation of the asymptotic expansion, the covariantly constant terms (and
some low-order derivatives terms), being summed up first. The covariant
algebraic approach is especially adequate and effective to study the low-energy
approximation. It seems that it can be developed deeper and that it can be
formulated a general technique for systematic calculation of the low-energy
heat kernel, a kind of {\it low-energy covariant perturbation theory}.

Among unsolved problems one should note, first of all, the problem of
generalizing our results to the most general covariantly constant background
including the Yang-Mills curvature.
Then, it is very interesting to obtain {\it explicitly} the covariantly
constant terms in HMDS-coefficients, i.e. to calculate $a_{k,k}$ part of
HMDS-coefficients.
This would be the opposite case to the higher-derivative approximation
and can be of certain interest in mathematical phy\-sics.
Finally, it is not perfectly clear how to do the analytical continuation to the
spacetime of Lorentzian signature.
All the activity in calculating the low-energy heat kernel is motivated by the
physical problem of studying the vacuum structure in quantum gravity and gauge
theories.

\bigskip
\bigskip
\leftline{\mittel Acknowledgments }
\bigskip

I would like to thank V. Berezin and V. Rubakov for their
kind invitation to present this talk at the International Seminar `Quantum
Gravity'. I am grateful to S. Fulling and many other participants of this
Seminar for many stimulating
and helpful discussions. I would like also to thank R. Schimming and J.
Eichhorn for their hospitality at the University of Greifswald. This work was
supported by the Alexander von Humboldt Foundation.

\bigskip
\bigskip

\leftline{\mittel References}
\bigskip

\item{[1]} J. S. Schwinger, Phys. Rev {\bf 82} (1951) 664
\item{[2]} B. S. De Witt,
	{\it Dynamical theory of groups and fields} (Gordon and Breach,
	New York, 1965)
\item{[3]} G. A. Vilkovisky,
	in: {\it Quantum theory of gravity}, ed.  S. Christensen (Hilger,
	Bristol, 1983) p. 169
\item{[4]} A. O. Barvinsky and G. A. Vilkovisky,
	Phys. Rep. {\bf C 119} (1985) 1
\item{[5]} I. G. Avramidi, {\it The covariant methods for calculation of
	the effective action in quantum field theory and the investigation
	of higher derivative quantum gravity}, PhD Thesis (Moscow State
	University, Moscow, 1987)
\item{[6]} I. G. Avramidi,
	Nucl. Phys. {\bf B 355} (1991) 712
\item{[7]} P. B. Gilkey,
	{\it Invariance theory, the heat equation and the  Atiyah - Singer
	index theorem} (Publish or Perish, Wilmington, DE, USA, 1984)
\item{[8]} J. Hadamard,
		{\it Lectures on Cauchy's Problem}, in: {\it Linear Partial
		Differential
		Equations}, (Yale U. P., New Haven, 1923)
\item{[9]} S. Minakshisundaram and A. Pleijel,
		Can. J. Math. {\bf 1} (1949) 242
\item{[10]} R. T. Seeley,
		Proc. Symp. Pure Math. {\bf 10} (1967) 288
\item{[11]} R. Schimming,
	Beitr. Anal. {\bf 15} (1981) 77
\item{[12]} R. Schimming, in {\it Analysis, Geometry and Groups: A Riemann
	Legacy Volume}, ed. H. M. Srivastava and Th. M. Rassias,
	(Hadronic Press, Palm Harbor, 1993), part. II, p. 627
\item{[13]} E. Elizalde, S. D. Odintsov, A. Romeo, A. A. Bytsenko and S.
Zerbini,
	{\it Zeta regularization techniques with applications} (World
	Scientific, Singapore, 1994)
\item{[14]} R.  Camporesi,
	Phys. Rep. {\bf 196} (1990) 1
\item{[15]} I. G. Avramidi,
	Yad. Fiz. {\bf 49} (1989) 1185
\item{[16]} I. G. Avramidi,
	Phys. Lett. {\bf B 236} (1990) 443
\item{[17]} A. O. Barvinsky, Yu. V. Gusev, G. A. Vilkovisky and
	V. V. Zhytnikov, J. Math. Phys. {\bf 35} (1994) 3543
\item{[18]} L. Parker and D. J. Toms, Phys. Rev. D {\bf 31} (1985) 953
\item{[19]} I. Jack and L. Parker, Phys. Rev. D {\bf 31} (1985) 2439
\item{[20]} J. A. Zuk, Phys. Rev. D {\bf 34} (1986) 1791
\item{[21]} J. A. Zuk, Phys. Rev.  D {\bf 33} (1986) 3645
\item{[22]} I. G. Avramidi,
	{\it Covariant methods for calculating the low-energy effective
	action in quantum field theory and quantum gravity},
	University of Greifswald (1994), gr-qc/9403036
\item{[23]} I. G. Avramidi,
	Phys. Lett. {\bf B 305} (1993) 27
\item{[24]} I. G. Avramidi, Phys. Lett. {\bf B 336} (1994) 171
\item{[25]} I. G. Avramidi,
	{\it A new algebraic approach for calculating the heat kernel
	in quantum gravity}, University of Greifswald (1994),
	hep-th/9406047, J. Math. Phys. (1995), to appear
\item{[26]} I. G. Avramidi, {\it New algebraic methods for calculating the heat
	kernel and the effective action in quantum gravity and gauge theories},
	University of Greifswald (1994),
	gr-qc/9408028; in {\it Heat Kernel Techniques and Quantum Gravity,
	Discourses in Mathematics and Its Applications},  No.~4, edited by
	S.~A. Fulling, Texas A\&M University, (College Station, Texas, 1995),
	to appear
\item{[27]} I. G. Avramidi, {\it Covariant algebraic calculation of the
	low-energy heat kernel}, University of Greifswald (1995),
	hep-th/9503132, J. Math. Phys. (1995), to appear
\item{[28]} I. G. Avramidi, J. Math. Phys. {\bf 36} (1995) 1557
\item{[29]} P. B. Gilkey,
	J. Diff. Geom. {\bf 10} (1975) 601
\item{[30]} I. G. Avramidi,
	Teor. Mat. Fiz. {\bf 79} (1989) 219
\item{[31]} I. G. Avramidi,
	Phys. Lett. {\bf B 238} (1990) 92
\item{[32]} P. Amsterdamski, A. L. Berkin and D. J. O'Connor,
	Class. Quantum  Grav. {\bf 6} (1989) 1981
\item{[33]} I. G. Avramidi and R. Schimming, {\it The heat kernel coefficients
	to the matrix Schr\"odinger operator}, University of Greifswald (1995),
	hep-th/9501026, J. Math. Phys. (1995), to appear
\item{[34]} T. Branson, P. B. Gilkey and B. \O rsted,
	Proc. Amer. Math. Soc. {\bf 109} (1990) 437
\item{[35]} J. A. Wolf, {\it Spaces of constant curvature},
	(University of California, Berkeley, CA, 1972)
\item{[36]} M. Takeuchi, {\it Lie Groups II, in Translations of Mathematical
Monographs}, vol. 85, (AMS, Providence, 1991)

\bye